\documentclass[10pt,twocolumn,aps,prd,nofootinbib]{revtex4-2}
\usepackage{amsmath,amssymb}  
\usepackage{graphicx}         
\usepackage{bm}               
\usepackage{epstopdf}
\usepackage[colorlinks=true,citecolor=red,linkcolor=blue,urlcolor=blue]{hyperref}

\begin{document}
\title{Recovering Semiclassical Einstein's Equation Using Generalized Entropy}
\author{Naman Kumar}
\affiliation{School of Basic Sciences, Indian Institute of Technology, Bhubaneswar, India, 752050}
\email{namankumar5954@gmail.com}

\begin{abstract}
    In this letter, we show that the Semiclassical Einstein's Field Equation can be recovered using the generalized entropy $S_{gen}$. This approach is reminiscent of non-equilibrium thermodynamics. Furthermore, contrary to the entanglement equilibrium approach of deriving the semiclassical Einstein's equation, this approach does not require any such assumptions and still recovers its correct form. Therefore, in a sense, we also show the validity of the semiclassical approximation, a crucial approach for establishing a number of important ideas such as the Hawking effect.
    \end{abstract}

\maketitle

\section{Introduction}
The connection between gravity and thermodynamics has long been explored. The first hints came from black hole thermodynamics when Hawking \cite{hawking1971gravitational} showed that the area of a horizon($A$) is a non-decreasing function of time.
\begin{equation}
    \frac{dA}{dt}\geq0
\end{equation}
Later, Bekenstein\cite{PhysRevD.7.2333} proposed the equality of horizon area and entropy as
\begin{equation}
    S=\gamma A
\end{equation}
where $\gamma$ is a constant of proportionality. Hawking \cite{1975CMaPh..43..199H} fixed this constant of proportionality by deriving the black hole temperature as\footnote{In this paper, we set $k_B=c=1$}
\begin{equation}
    \gamma=\frac{1}{4G\hbar}
\end{equation}
where $G$ is the universal gravitational constant and $\hbar$ is the reduced Planck's constant. In 1995, Jacobson \cite{Jacobson1995ThermodynamicsOS} derived Einstein's field equation from this proportionality of horizon area and entropy and the fundamental thermodynamic result $dQ=TdS$ under equilibrium conditions. Some works under the non-equilibrium conditions were also presented later \cite{eling2006nonequilibrium,chirco2010nonequilibrium}. Jacobson took $Q$ as the heat flux across a causal horizon while temperature $T$ was given by Unruh temperature\cite{unruh1976notes} as observed by a local Rindler observer. Using these ingredients, he was able to derive Einstein's field equation in the form
\begin{equation}
    R_{ab}-\frac{1}{2}Rg_{ab}+\Lambda g_{ab}=\frac{2\pi}{\hbar\eta}T_{ab}\label{jacobfe}
\end{equation}
where the constant $\eta=(4G\hbar)^{-1}$ and the constant $\Lambda$ was identified as the cosmological constant. The connection between gravity and thermodynamics was further explored by Padmanabhan (see \cite{padmanabhan2010thermodynamical} for a review) with some closely related follow-up articles from his collaborators \cite{mukhopadhyay2006holography,kothawala2007einstein,kothawala2009thermodynamic,kolekar2010holography}. Recently, Verlinde\cite{verlinde2011origin} combined thermodynamic arguments and the holographic principle\cite{susskind1995world,hooft1993dimensional} and argued that gravity is an entropic force arising from the underlying microscopic theory to maximize its entropy. Very recently, in 2016, Bousso et al.\cite{bousso2016quantum} conjectured the Quantum Focusing(QFC), which conjectures that the quantum expansion($\Theta$) cannot increase along any congruence, even in quantum states that would violate the classical focusing theorem:
\begin{equation}
    \frac{d\Theta}{d\lambda}\leq0
\end{equation}
The quantum expansion allows us to generalize the classical focusing theorem to the semiclassical case. The QFC has a wide variety of applications. It implies a Quantum Bousso Bound which has already gathered a considerable amount of evidence \cite{bousso1999covariant,bousso2002holographic,bousso2003simple,bousso2014proof,bousso2015entropy,bousso2016quantum,flanagan2000proof,strominger2004quantum}. It also implies the quantum singularity theorems \cite{wall2013generalized,bousso2022quantum}, the generalized second law of causal horizons and holographic screens \cite{bousso2016generalized} and new property of nongravitational theories, the Quantum Null Energy Condition(QNEC)\cite{bousso2016quantum,bousso2016proof,balakrishnan2019general,ceyhan2020recovering}. Arvin \cite{shahbazi2022restricted} presented a restricted quantum focusing which he argued is sufficient to derive all known essential implications of the quantum focusing and also proved it on brane-world semiclassical gravity theories. The Semiclassical Einstein's Field equation was derived using the entanglement equilibrium approach \cite{jacobson2016entanglement,bueno2017entanglement} which was essentially based on two assumptions: first, the vacuum reduced to a small ball is maximal when we vary it keeping the volume fixed and second, the variation of the entanglement entropy coming from the variation of the matter fields takes the following particular form
\begin{equation}
   \delta S_{IR}=2\pi\frac{\Omega_{d-2}R^d}{d^2-1}\bigg(\delta\langle T_{00}\rangle-\delta\langle X\rangle\bigg) 
\end{equation}
where $X$ is an operator in the QFT and the remaining symbols have the usual meaning. This second assumption was critically analyzed and was found to be somewhat problematic to an extent such that the entanglement equilibrium approach can only reveal linearized Einstein's equation or the full non-linear equation by restricting to some special choices of states(see \cite{casini2016comments} for details).
In this letter, we show that the generalized entropy approach can bypass all such assumptions and we can recover the full non-linear Semiclassical Einstein's Field equation, which reflects on the naturalness and robustness of our approach. In the next section, we present our idea elaborately.

\section{Recovering the Semiclassical Einstein's Field Equation}
\begin{figure}[ht]
    \centering
    \includegraphics[width=0.5\textwidth]{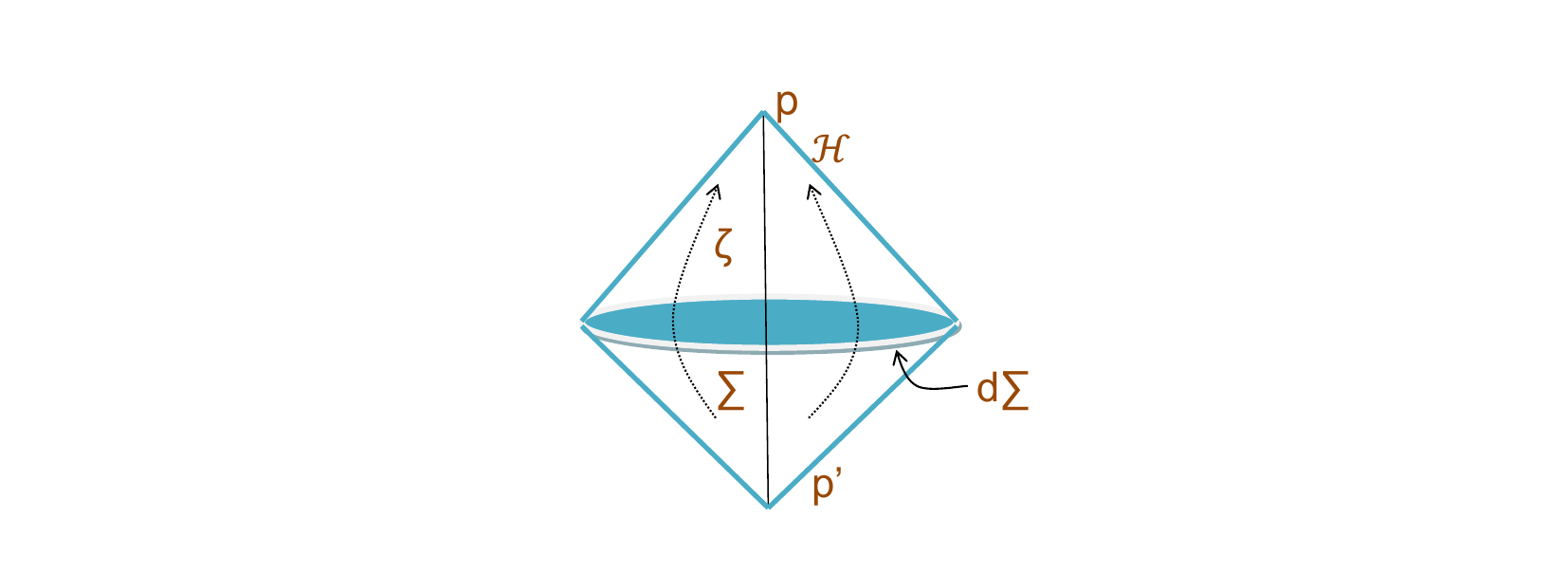}
    \caption{A causal diamond in the de-Sitter static patch for a ball-shaped spacelike surface $\Sigma$ with $d\Sigma$ as its boundary. $p$ and $p'$ are the future and past vertices of the diamond. $\zeta$ is a timelike killing vector in the large limit of the diamond when the boundary of the diamond coincides with the cosmological horizon. The dashed curves are the flow lines of the killing vector $\zeta$. $\mathcal{H} $ is the killing horizon.}
    \label{fig1}
\end{figure}

Hawking \cite{1975CMaPh..43..199H} showed that black holes could evaporate, thus their horizon area can decrease via Hawking radiation. One of the key assumptions he made is the validity of semiclassical Einstein's field equation in the semiclassical regime given by
\begin{equation}
    G_{\mu\nu}=\frac{8\pi G}{c^4}\langle T_{\mu\nu}\rangle\label{semfe}
\end{equation}
where $G_{\mu\nu}$ is the Einstein tensor and $\langle T_{\mu\nu}\rangle$ is the expectation value of the energy-momentum tensor. We show how to recover this critical equation, thus establishing the validity of the semiclassical approximation under the assumption that the fluctuations in $T_{ab}$ are negligible.\\ We consider a causal diamond in the large limit in the de-Sitter static patch (see Fig.(\ref{fig1})). In this limit, the boundary of the diamond coincides with the cosmological horizon of the dS space\footnote{The metric for a static patch in de-Sitter space is given by $ds^2=-[1-(r/L)^2]dt^2+[1-(r/L)^2]^{-1}dr^2+r^2d\Omega^2_2$}, that is, $r=L$. The conformal killing vector $\zeta^a$ then becomes the timelike killing vector and $\mathcal{H}$ the killing horizon\cite{jacobson2019gravitational}.
\\
The generalized entropy $S_{gen}$ is defined as
\begin{equation}
S_{gen}[\mathcal{H}]=\frac{A[\mathcal{H}]}{4G\hbar}+S_{out}[\mathcal{H}]    
\end{equation}
where $S_{out}[\mathcal{H}]$ is the entropy restricted to one side of $\mathcal{H}$ (here, the matter entropy outside the horizon $\mathcal{H}$ of the causal diamond). The Quantum Expansion ($\Theta$) is defined similarly to the classical expansion $\theta$ as the response of $S_{gen}$ to deformations of $\mathcal{H}$ along the generator and the amount of deformation is measured by an affine parameter $\lambda$ along the generator. This translates to the definition of $\Theta$ as
\begin{equation}
 \Theta=\lim_{\mathcal{A}\to0}
\frac{4G\hbar}{\mathcal{A}}\frac{dS_{gen}}{d\lambda}
\end{equation}
Therefore, the variation of the generalized entropy is given by
\begin{equation}
    \delta S_{gen}=\eta\int_{\mathcal{H}}\Theta d\lambda d\mathcal{A}\label{delsgen1}
\end{equation}
where $\eta=(4G\hbar)^{-1}$. The cosmological horizon $\mathcal{H}$ has a temperature given by
\begin{equation}
    T=\frac{\hbar\kappa}{2\pi}
\end{equation}
where $\kappa$ is the surface gravity. Since the interior of the causal diamond is unobservable as it is causally disconnected from the exterior, the integrated energy of the energy current of the matter given by $\langle T_{ab}\rangle\zeta^a$, is identified as heat\cite{Jacobson1995ThermodynamicsOS}. 
The energy flux across $\mathcal{H}$ can therefore be written as
\begin{align}
\delta Q=\int_{\mathcal{H}}d\Sigma^a\langle T_{ab}\rangle \zeta^b\\
    =-2\kappa\int_{\mathcal{H}}\lambda \langle T_{ab}\rangle k^ak^bd\lambda d\mathcal{A}\label{heat}
\end{align}
where $k^a$ is the (null) tangent vector to the horizon generators for an affine parameter $\lambda$ so that $\zeta^a=-2\kappa\lambda k^a$. Note the extra factor of $2$ in the equation of heat (\ref{heat}). We recall that the relation between the affine parameter $\lambda$ and the killing parameter $v$ on a killing horizon is generally given as $\lambda=-e^{-\kappa v}$. But as we show here, in the semiclassical case, we should use instead $\lambda=-e^{-2\kappa v}$ so as to recover the semiclassical equation. This can be understood as the vanishing of quantum expansion $\Theta$ to zeroth order in $\lambda$ must occur at twice the rate to correctly impose the condition on spacetime curvature.
Therefore, we get $\zeta^a=-2\kappa\lambda k^a$, and $\delta Q$ takes the form as given in (\ref{heat}). In the semiclassical case, the natural generalization to work with the generalized entropy is quantum expansion. The infinitesimal evolution of $\Theta$ is given by a linear expansion around its equilibrium value at $p$, up to the first order in $\lambda$, as it moves away from the equilibrium surface at $\lambda=0$ along the generator 
\begin{equation}
\Theta=\Theta_p+\lambda\frac{d\Theta}{d\lambda}\bigg|_p+\mathcal{O}(\lambda^2)    
\end{equation}
To find $\Theta$ in terms of known quantities, we use its definition as \cite{bousso2016quantum}
\begin{equation}
\Theta=\theta+\frac{4G\hbar}{\mathcal{A}}S'_{out}
    \end{equation}
where $\theta$ is the classical expansion and $S_{out}$ the outside matter entropy. The evolution of Quantum Expansion $\Theta$ is governed by
\begin{align}
    \frac{d\Theta}{d\lambda}=\frac{d\theta}{d\lambda}
    +\frac{d}{d\lambda}\bigg(\frac{4G\hbar}{\mathcal{A}}S'_{out}\bigg)\\
    =-\frac{1}{2}\theta^2-\xi^2-R_{ab}k^ak^b+\frac{d}{d\lambda}\bigg(\frac{4G\hbar}{\mathcal{A}}S'_{out}\bigg)\label{dThetadlambda}
\end{align}
Therefore (\ref{delsgen1}) becomes
\begin{equation}
\begin{split}
\delta S_{gen}=\eta\int_{\mathcal{H}}\bigg[\Theta_p-\lambda\bigg(\frac{\theta^2}{2}+\xi^2+R_{ab}k^ak^b\\-\frac{d}{d\lambda}\frac{4G\hbar}{\mathcal{A}}S'_{out}\bigg)\bigg|_p\bigg]d\lambda d\mathcal{A}\label{delsgen}
    \end{split}
\end{equation}
Since at zero order\footnote{As can be noted here. In contrast with the classical case, the semiclassical case requires the vanishing of quantum expansion $\Theta$ to zeroth order and not the classical expansion $\theta$ for the semiclassical Einstein's equation to hold. This can be understood as follows: The quantum fields violate the classical focusing, while even in this, the quantum focusing ($d\Theta/d\lambda\leq0$) holds. Therefore, focusing to the past of $p$ must bring the quantum expansion to zero so that the increase in generalized entropy $S_{gen}$ is proportional to the killing energy across it. This imposes a condition on the spacetime curvature that is governed by the evolution equation (\ref{dThetadlambda}). The same cannot be said for the classical focusing since it does not hold in the semiclassical case.} in $\lambda$, $\Theta_p=0$, we have
\begin{equation}
    \theta_p=-\frac{4G\hbar}{\mathcal{A}}S'_{out}\bigg|_p
\end{equation}
We therefore get
\begin{equation}
\delta S_{gen}=\eta\int_{\mathcal{H}}\bigg[-2\lambda\bigg(\frac{\theta^2}{2}+\xi^2+R_{ab}k^ak^b\bigg)\bigg]d\lambda d\mathcal{A}\label{delsgen2}
\end{equation}
This equation follows from the fact that:
In the regime of small perturbations about the Minkowski vacuum, one may treat the causal horizon as an (approximately) Rindler horizon (the near-horizon geometry for the de-Sitter static patch that we are working with is also Rindler), for which the \emph{first law of entanglement} holds off the bifurcation surface. Under this near-equilibrium, Rindler-horizon approximation, we \emph{assume}
\begin{equation}
    \delta S_{\rm out}=\frac{\delta A}{4G}\label{delta_S_out}
\end{equation}
This relation is justified by the perturbative quantum gravity argument of Bianchi \& Satz\footnote{See \cite{bianchi2012black,bianchi2013mechanical}, where for an infinitesimal perturbation of the Minkowski vacuum one shows
\[
\Delta S_{\rm ent}
=\frac{\Delta A}{4G}
\]
to leading order in the perturbation.}
, which proves that for any small energy–momentum flux through a Rindler horizon
\begin{equation}
\begin{split}
    \delta S_{\rm out}
=\delta\langle K\rangle
=2\pi\!\int_H(-\lambda)\,T_{ab}k^ak^b\,d\lambda\,dA,
\quad\\
\delta A
=-8\pi G\!\int_H(-\lambda)\,T_{ab}k^ak^b\,d\lambda\,dA, 
\end{split}   
\end{equation}
Hence, generator by generator
\begin{equation}
    \delta S_{\rm out}
=\frac{\delta A}{4G}
\end{equation}
Therefore, under the assumption (\ref{delta_S_out}), the step from (18) to (20) follows immediately, since 
\begin{equation}
S'_{\rm out}(\lambda)
=\frac{\delta S_{\rm out}}{\delta\lambda}
=\frac{1}{4G}\,\frac{\delta A}{\delta\lambda}
=\frac{A(\lambda)}{4G}\,\theta(\lambda).
\end{equation}
which, on differentiating, gives
\begin{equation}
    \frac{d}{d\lambda}\!\Bigl(\frac{4G}{A}\,S'_{\rm out}\Bigr)
=\frac{d\theta}{d\lambda}
\end{equation}
We now invoke the Clausius relation $\delta Q=T\delta S$ but before we do that, let us write $\delta S_{gen}$ as $\delta S_{gen}=d_iS+d_eS$, where
\begin{equation}
    d_iS=-2\eta\int_{\mathcal{H}}\lambda\bigg(\frac{\theta^2}{2}+\xi^2\bigg)d\lambda d\mathcal{A}\label{dis}
\end{equation}
and
\begin{equation}
    d_eS=-2\eta\int_{\mathcal{H}}\lambda R_{ab}k^ak^bd\lambda d\mathcal{A}\label{des}
\end{equation}
$d_iS$ is identified as the internal entropy production rate due to some dissipation \cite{chirco2010nonequilibrium,dey2017spacetime}, and it must be positive as required by the second law. Interestingly, this is twice that of the classical one and has contributions from both scalar and tensorial degrees of freedom. To see why $d_eS$ and $d_iS$ take these expressions can be understood by expressing (\ref{dis}) in terms of the Killing parameter $v$ as
\begin{equation}
    d_iS=\frac{\eta}{\kappa}\int_{\mathcal{H}}dvd\mathcal{A}\bigg(\frac{\theta^2}{2}+\xi^2\bigg)\geq0
\end{equation}
as required by the second law (since the affine parameter $\lambda$ and Killing parameter $v$ on the horizon are related by $\lambda=-e^{-2\kappa v}$). $d_eS$ is then identified as the reversible entropy change $dS_{rev}$. Therefore, by subtracting $d_iS$ from $\delta S_{gen}$ and using the Clausius relation $\delta Q=TdS_{rev}=Td_eS$, we obtain
\begin{equation}
    \langle T_{ab}\rangle k^ak^b=\frac{\hbar\eta}{2\pi}R_{ab}k^ak^b
\label{maineq}
\end{equation}
This equation implies
\begin{equation}
    \frac{2\pi}{\hbar\eta}\langle T_{ab}\rangle=R_{ab}+fg_{ab}
\end{equation}
for some function $f$. Local conservation implies $\langle T_{ab}\rangle$ is divergenceless; therefore, using the Bianchi identity, we recover the semiclassical Einstein's equation
\begin{equation}
    R_{ab}-\frac{1}{2}Rg_{ab}+\Lambda g_{ab}=\frac{2\pi}{\hbar\eta}\langle T_{ab}\rangle
\end{equation}
where $\Lambda$ is identified as the cosmological constant of the de-Sitter space and $\eta=(4G\hbar)^{-1}$. Therefore, the constant $\Lambda$ obtained in this case also carries a natural meaning.
\section{Conclusion}
In this work, we showed that the semiclassical Einstein's field equation can be recovered in the semiclassical case using the generalized entropy $S_{gen}$. Since $S_{gen}$ is a cutoff independent quantity, it reveals information about the full quantum gravity theory and shows that the semiclassical approximation can be trusted as long as the fluctuations in $T_{ab}$ are negligible. Furthermore, we showed that this approach is reminiscent of non-equilibrium thermodynamics. The validity of the semiclassical equation has been previously established using the entanglement equilibrium approach. However, this approach applied to causal diamonds makes a number of assumptions and some of which turned out to be problematic. On the contrary, we need no such assumptions, and still, we were successful in recovering the semiclassical equation, making this approach natural and robust.
\section*{Acknowledgements}
We would like to thank the anonymous referee who provided detailed comments and useful suggestions that greatly improved the quality of the manuscript.

\bibliographystyle{unsrt}
\bibliography{references}

\end{document}